\documentclass[useAMS]{gGAF2e}

\def\gsim{\lower.4ex\hbox{$\;\buildrel >\over{\scriptstyle\sim}\;$}}
\def\lsim{\lower.4ex\hbox{$\;\buildrel <\over{\scriptstyle\sim}\;$}}
\begin{document}
\doi{10.1080/03091920xxxxxxxxx}
\issn{1029-0419} \issnp{0309-1929} \jvol{00} \jnum{00} \jyear{2009} 

\markboth{G. R\"udiger and L.L. Kitchatinov }{Kink-type instability of toroidal stellar magnetic fields}

\title{{\textit{
The kink-type  instability of toroidal stellar magnetic fields with thermal diffusion$^{*}$\footnote{$^*$dedicated to Andrew Soward on the occasion of his 65$^{\rm th}$ birthday}
}}}

\author{G. R\"udiger$^1$$^{\ast\ast}$ and L.L. Kitchatinov$^{1,2,3}$
\\\vspace{6pt}
$^1$ Astrophysikalisches Institut Potsdam, An der Sternwarte 16, D-14482, Potsdam, Germany\\
$^2$\thanks{$^{\ast\ast}$Corresponding author. Email:
GRuediger@aip.de \vspace{6pt}} Institute for Solar-Terrestrial
Physics, PO Box 291, Irkutsk, 664033, Russia\\
$^3$ Pulkovo Astronomical Observatory, St. Petersburg, 196140,
Russia
\\\vspace{6pt}\received{v3.3 released February 2009} }

\maketitle

\begin{abstract}
The stability of toroidal magnetic fields in rotating radiative stellar zones is studied for realistic  values of both  the Prandtl numbers. The two considered  models for  the magnetic geometry represent fields with odd  and even symmetry with respect to the equator.  In  the linear theory in Boussinesq approximation the resulting complex eigenfrequency (including growth rate and drift  rate) are calculated for a  given radial wave number of a nonaxisymmetric perturbation with $m=1$.  The ratio of the Alfv\'en frequency, $\Omega_{\rm A}$, to the rate  of the basic rotation, $\Omega$,  controls the eigenfrequency of the solution. For strong fields with $\Omega_{\rm A}>\Omega$ the solutions do not feel the thermal diffusion.  The growth rate runs with $\Omega_{\rm A}$ and the drift rate is close to $-\Omega$ so that the magnetic pattern will rest in the laboratory system. For weaker fields with $\Omega_{\rm A}<\Omega$  the growth rate strongly depends on the thermal conductivity. For fields with dipolar parity and for typical  values of the heat conductivity the resulting very small growth rates are almost identical with those for   vanishing gravity. For fields with dipolar symmetry the differential  rotation of any stellar radiative zone (like the solar tachocline) is shown as basically stabilizing the instability independent of the sign of the shear.

Finally,  the current-driven  kink-type instability of a toroidal background field is proposed as a model for  the magnetism  of Ap stars. The recent observation of a lower magnetic field  treshold of about 300 Gauss for Ap stars is understood as corresponding to the minimum magnetic field producing the instability.
 \bigskip
 \begin{keywords}
 {stellar magnetic fields;  stellar rotation; differential rotation; solar tachocline}
 \end{keywords} \bigskip
 \end{abstract}

\section{Introduction}
This paper considers the stability of toroidal magnetic fields in
rotating radiation zones of stars and focuses on the destabilizing
effect of finite thermal diffusion and the stabilizing effect of
differential rotation. The equations for the linear stability of
toroidal magnetic fields under the influences of basic rotation and
gravitational buoyancy are  solved for two different latitudinal
profiles of the toroidal field, i.e. with the two possible
symmetries with respect to the equator. Our equations are global in
both horizontal dimensions but they  are local in radius, i.e. the
short-wave approximation  $kr \gg 1$  is used ($k$ is the radial
wave number).

\begin{figure}[h]
\begin{center}{
\resizebox*{7.1cm}{!}{\includegraphics{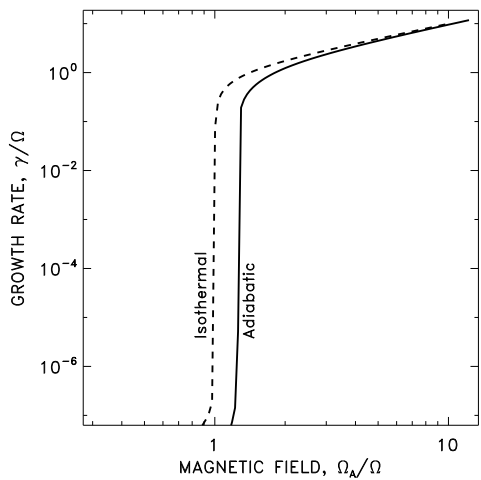}}}
\caption{\label{f0} Normalized growth rates of the nonaxisymmetric ($m=1$) kink-type instability    of an equatorsymmetric toroidal field in a sphere with $\rm Pm=5\times      10^{-3}$. The solid line represents the adiabatic approximation with $\rm   \chi=0$. The dashed line gives the limit of $\chi\to \infty$. The stabilizing   effect of buoyancy is rather small.}
\end{center}
\end{figure}

We know that a toroidal magnetic field $B_\phi$ which fulfills the condition
\begin{equation}
\frac{{\rm d}}{{\rm d}R} \left(R B_\phi^2\right) > 0
\label{eq1}
\end{equation}
is unstable  against nonaxisymmetric disturbances in an ideal and
incompressible medium (Tayler 1957, 1973, Vandakurov 1972, Acheson
1978). Here $R$ is the distance from the axis where $B_\phi\equiv
0$. This `Tayler instability' (TI) is suppressed by rigid rotation
unless the magnetic field is strong enough to fulfill the condition
\begin{equation}
\Omega_{\rm A}\geq \Omega
\label{eq2}
\end{equation}
(Pitts and Tayler 1985) with the magnetic frequency $\Omega_{\rm
A}=B_\phi/\sqrt{\mu_0\rho} R$  and the rotational frequency
$\Omega=\rm const$ of the star. For a simple spherical model with
$\Omega_{\rm A}=$ const (one magnetic belt with maximum in the
equatorial plane) the dashed line in figure~\ref{f0} demonstrates
this situation where the solar value of the magnetic Prandtl number
${\rm Pm}=\nu/\eta$ of $5\times 10^{-3}$ was used. Here $\nu$ and
$\eta$ are the microscopic values of the viscosity and magnetic
diffusivity. The growth rate $\gamma$ at the vertical axis of the
plot is normalized with the basic rotation so that the  result of
the calculation is
\begin{equation}
\gamma \propto \Omega_{\rm A} \quad\quad {\rm for} \quad\quad \Omega_{\rm A}\gsim \Omega .
\label{eq3}
\end{equation}
The instability thus  exists only for subAlfv\'enic rotation and it
is obviously very fast, i.e. $\gamma \simeq  \Omega_{\rm A}>
\Omega$.

Due to the assumed isothermal state of the medium no buoyancy-term
exists  in this calculation. Generally, in stably-stratified plasma
the real buoyancy is `negative' and should stabilize the magnetic
instability. This is indeed the case. The solid line in
figure~\ref{f0} results from a model with adiabatic density
fluctuations in Boussinesq approximation, i.e.  for vanishing
thermal diffusivity $\chi$. The resulting field strength for onset
of instability slightly exceeds its value without buoyancy.
Surprisingly, the stabilization of the TI by `negative buoyancy' is
a small effect. The opposite case with  $\chi\to \infty$ provides
results identical with those  for the incompressible model (dashed
line). No temperature fluctuations can develop  for  $\chi \to
\infty$ so that the stabilizing effect of buoyancy does not apply.

After  inspection of the extremes $\chi \to 0$ and $\chi \to \infty$  one could believe that the inclusion of the thermal equation into the instability theory is not significant. We know, however, that the heat conductivity  in the stellar radiation zones is the most effective dissipation process, far from the adiabatic limit, i.e.
\begin{equation}
\chi \gg \eta \gg \nu .
\label{eq4}
\end{equation}
The Prandtl number  of the gas is of order
\begin{equation}
{\rm Pr}= \frac{\nu}{\chi}\simeq 2\times 10^{-6}
\label{eq5}
\end{equation}
which yields  a Roberts number ${\rm q}=\chi/\eta$ of ${\rm q}\simeq 2500$. Note that liquid
metals of MHD laboratory experiments have ${\rm q}\ll 1$.

The nonadiabatic models with small Prandtl number (see equation (\ref{eq5})) will lead to a surprising result. The buoyancy does {\em not} suppress the instability any longer but it creates another instability  which exists for much weaker magnetic fields. Acheson (1978) found that the Roberts number q enters the stability  equations in a rather complicated manner so that  one term acts stabilizing while another one acts destabilizing. ``It is natural, therefore, to expect quite complicated changes in the stability properties  of the system ...''. All terms, however, vanish when  ${\rm q}\to \infty$. We shall find  that indeed finite thermal conductivities lead to a  destabilization of magnetic fields  with  $\Omega_{\rm A}$ much smaller than $\Omega$;  the growth rates of this doubly-diffusive  instability, however,   are small. Neither   time stepping codes nor  technical experiments can find such slow instabilities with growth rates of order  $10^{-4}\Omega$. Of course, such timescales are still very short compared to stellar evolutionary  times but the physical relevance of such instabilities is still in question (Cally 2003).

Critical for the astrophysical meaning of nonaxisymmetric magnetic instabilities is also the differential rotation.
One expects that the nonuniform rotation acts against the excitation of nonaxisymmetric modes if they  are
already excited for rigid rotation. In the solar tachocline the rotation law is -- per definition -- nonuniform. The differential rotation itself  is  needed for the production of the  strong toroidal fields leading to   the tachocline phenomenon. A possible magnetic instability of these toroidal
fields (``Tayler-Spruit dynamo'') should  basically be  stabilized by the nonuniformity of the rotation. One has thus to check how important the  differential rotation is to stabilize/destabilize the toroidal magnetic fields. There are  indications that the influence of the differential rotation strongly depends on the  symmetry type of the toroidal field patterns with respect to the equator.
In the present paper only fields with antisymmetry with respect to the equator are considered as subject to differential rotation.
\section{The model}
The equations for small disturbances ($'$) of background magnetic ($\bar{\bf B}$) and velocity ($\bar{\bf u}$) fields are
\begin{eqnarray}
 \frac{\partial{\bf u}'}{\partial t}
 &+& \left(\bar{\bf{u}}\cdot\nabla\right){\bf u}'
 + \left({\bf u}'\cdot\nabla\right)\bar{\bf{u}}
 + \frac{1}{{\mu_0\rho}}\left(\nabla\left(\bar{\bf{B}}\cdot{\bf B}'\right)-\right.
 \nonumber \\
 &-& \left.\left(\bar{\bf{B}}\cdot\nabla\right){\bf B}'
 - \left({\bf B}'\cdot\nabla\right)\bar{\bf{B}}\right)
 =-\left(\frac{1}{\rho}\nabla P\right)' + \nu\Delta{\bf u}'
  \label{5}
\end{eqnarray}
for the velocity fluctuations $\bf{u}'$,
\begin{equation}
   \frac{\partial{\bf B}'}{\partial t} =
   {\rm curl} \left(\bar{\bf{u}}\times{\bf B}'
   + {\bf u}'\times\bar{\bf{B}} - \eta\ {\rm curl}{\bf B}'\right) ,
   \label{6}
\end{equation}
for the magnetic fluctuations $\bf{B}'$ and
\begin{equation}
  \frac{\partial s'}{\partial t} + \bar{\bf{u}}\cdot \nabla s' +
  {\bf u}'\cdot\nabla \bar s = \frac{C_\mathrm{p}\chi}{T}\Delta T'
  \label{7}
\end{equation}
for the entropy fluctuations $s'$ related to the density fluctuations by $s' =-C_\mathrm{p}\rho'/\rho$.

The basic flow is a rotation with uniform angular velocity $\Omega$ and the mean magnetic field $\bar{\bf{B}}$ has only a toroidal component $B_\phi$.
The mathematical method for solving the equation system (\ref{5})--(\ref{7}) has been described earlier (Kitchatinov and R\"udiger 2008). In this previous paper the instability has been considered for the case of rigid rotation and for given heat conductivity. In the present  paper the {\em destabilizing} role of the heat conductivity is demonstrated and first results for the {\em stabilizing} role of differential rotation are given.

The equations are global in horizontal dimensions but local in
radius. The radial scale of disturbances is assumed as short and
their dependence on radius $r$ is taken in the form of  Fourier
modes $\mathrm{exp}(\mathrm{i}(m\phi-\omega t + kr))$. There are
only symmetry conditions along  the polar axes  which will be
fulfilled by the series expansions after Legendre polynomials. Only
the modes with $m=1$ are considered. Then at the axes the radial
components of  flow and field must vanish and also the
$\theta$-derivatives of the horizontal components (see Elstner {\it
et al.} 1990, Gilman and Fox 1997).  For given radial wave number
and for given field amplitude (in units of the basic rotation
velocity) the resulting eigenfrequency is computed including growth
rate  (imaginary part of $\omega$) of the instability and drift rate
of the eigensolutions (real part of $\omega$).

The key
parameter for the effect of the stable stratification
is
 \begin{equation}
    \hat\lambda = \frac{N}{\Omega kr} ,
    \label{1}
 \end{equation}
where $N$ is the buoyancy frequency
\begin{equation}
  N^2= \frac{g}{C_\mathrm{p}}\frac{\partial s}{\partial r}.
    \label{1a}
 \end{equation}
In stellar radiation zones it is $N \gg \Omega$. For the most unstable modes we  find $\hat\lambda < 1$
so that the radial scale of the modes is indeed small, i.e. $kr \gg 1$.
Our equations include finite diffusion via
\begin{equation}
    \epsilon_\eta = \frac{\eta N^2}{\Omega^3 r^2},\ \ \ \ \ \ \ \ \ \ \ \ \ \ \ \ \ \ \ \ \
    \epsilon_\nu = \frac{\nu N^2}{\Omega^3 r^2},
    \label{2}
\end{equation}
  with $\eta$ and $\nu$ as the magnetic resistivity and the kinematic viscosity.
The thermal conductivity $\chi$ enters the equations in the normalized form
\begin{equation}
    \epsilon_\chi = \frac{\chi N^2}{\Omega^3 r^2},
    \label{21}
\end{equation}
which is the free parameter in the following discussion.

The present  article focuses on the effect of thermal diffusivity
from the following reason. The current-driven Tayler instability
requires radial displacements. The instability does not exist in the
2D case of purely horizontal disturbances (see Dicke 1979). In a
stably stratified radiation zone the radial displacements are
opposed by buoyancy. Finite thermal diffusivity reduces the buoyancy
and thus supports the instability. The relevant parameter is the
ratio $C_\chi$
\begin{equation}
    C_\chi = \frac{\chi k^3 r}{N} = \frac{\epsilon_\chi}{{\hat{\lambda}}^3}
    \label{3}
 \end{equation}
of the
frequency $\chi k^2$, with which the thermal diffusion destroys the
buoyancy, to the characteristic frequency $N/(kr)$ of gravity waves.
 When $\epsilon_\chi$ and $\hat\lambda$ are simultaneously varied  for constant
 $C_\chi$-parameter (\ref{3}) the
results change little.

The toroidal field profile of the model is parameterized as
 \begin{equation}
    B_\phi = \sqrt{\mu_0\rho}\ r\sin\theta\cos^n\theta\ \Omega_\mathrm{A},
    \label{4}
 \end{equation}
where  now  $\Omega_\mathrm{A}$ is  the {\em amplitude} of the
magnetic Alfv\'en frequency. Computations  were made for  $n=0$ so
that the toroidal field is symmetric (or quadrupolar) with respect
to the equator. The other model with $n=1$  represents a (dipolar)
background field antisymmetric relative to the equator or -- in
other words -- with two belts of opposite signs.

The diffusion parameters $\epsilon_\eta = 4\times 10^{-8}$ and
$\epsilon_\nu = 2\times 10^{-10}$ characteristic for the upper
radiation zone of the Sun are kept fixed and $\epsilon_\chi$ is
varied to study the effect of thermal diffusion.

The normalized wavelengths (\ref{1}) were also kept constant. The
constant values were $\hat\lambda = 0.6$ for quadrupolar background
field ($n=0$)and $\hat\lambda = 0.1$ for dipolar fields ($n=1$). The
choice is motivated by the finding that for this values of
$\hat\lambda$ the marginal field strengths for onset of the
instability are minimized. The
values of $\hat\lambda$ corresponding to maximum growth rates of
supercritical excitations change with the external parameters but
only slightly. Some results of this paper were obtained for the
instability modes of symmetry type S1, the equatorially symmetric
excitations with azimuthal wave number $m=1$. For $\hat\lambda =
0.1$ and for the given Prandtl number (\ref{eq5}) one finds
$C_\chi=0.1$ as  the solar value. Computing the critical magnetic frequencies under the presence of differential rotation the  wave numbers have been varied as long as the minimum eigenvalues have been found (see below).

The linear code is able to calculate the eigenfrequencies for
the realistic small magnetic Prandtl number
 \begin{equation}
    \rm Pm = 5 \times 10^{-3}.
    \label{41}
 \end{equation}
So far the best nonlinear MHD codes  reach values down to 0.001 \citep{Cea05, B09}.
\section{Results}
In the following the stability of the  magnetic background field (\ref{4}) is probed for nonaxisymmetric  disturbances with the azimuthal wave number $m=1$ and with a fixed radial scale. By use of the method described by Kitchatinov and R\"udiger (2008) the  complex equation system (\ref{5})--(\ref{7}) is numerically solved to find the  growth rate  of a possible instability. Generally, the results do not change if $m$ is replaced by $-m$. Indeed, for purely toroidal magnetic fields there is no handedness in the system. The addition of even  a weak poloidal field  would break the symmetry between $m$ and $-m$.

\subsection{Quadrupolar magnetic geometry}
We start with $n=0$ as the most simple case. As the fields are symmetric with respect to the equator this model resembles stellar models with toroidal fields of quadrupolar parity.

\begin{figure}
\begin{center}{
\resizebox*{7cm}{!}{\includegraphics{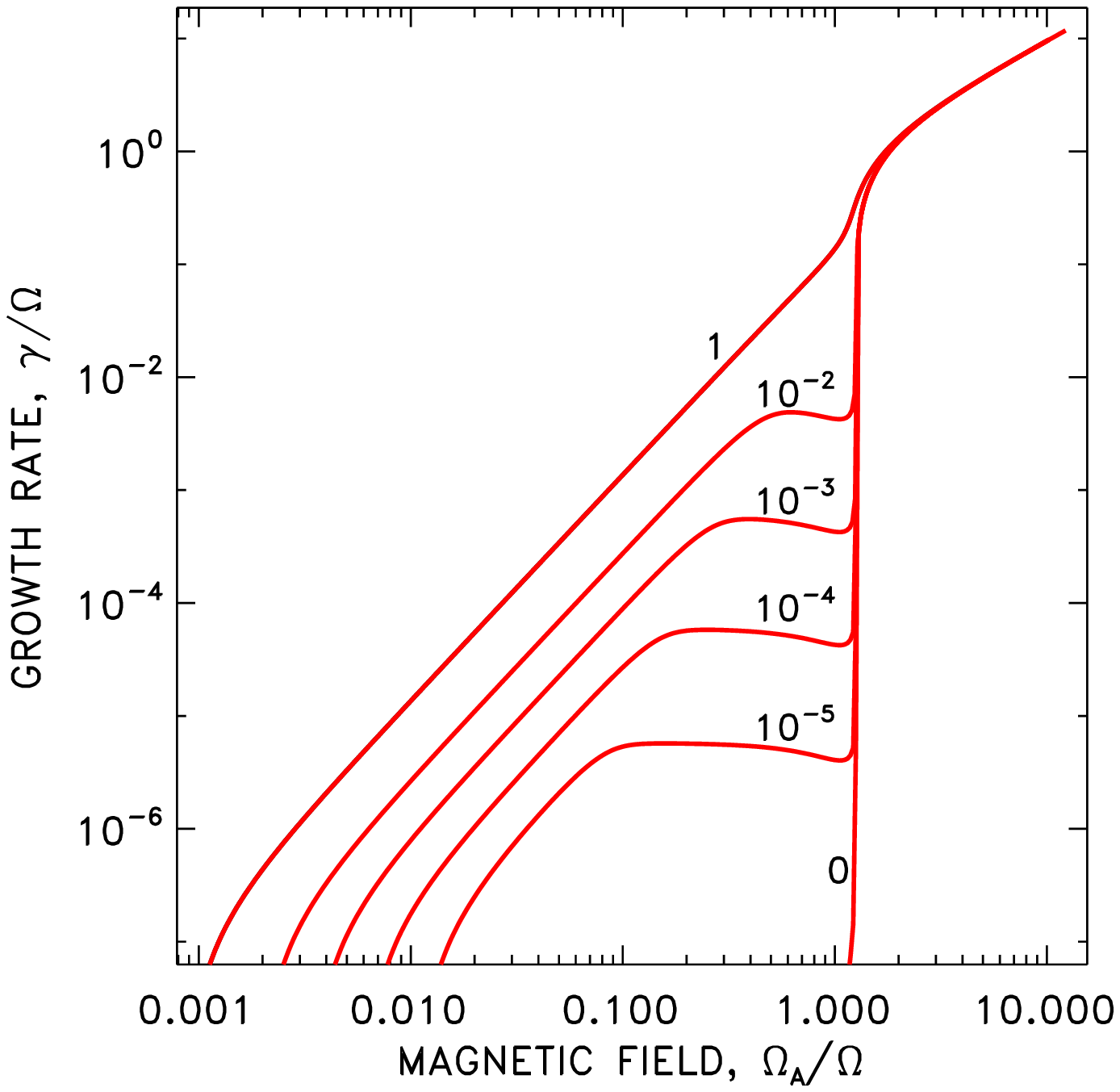}}
\hspace{0.3truecm}
\resizebox*{7.1cm}{!}{\includegraphics{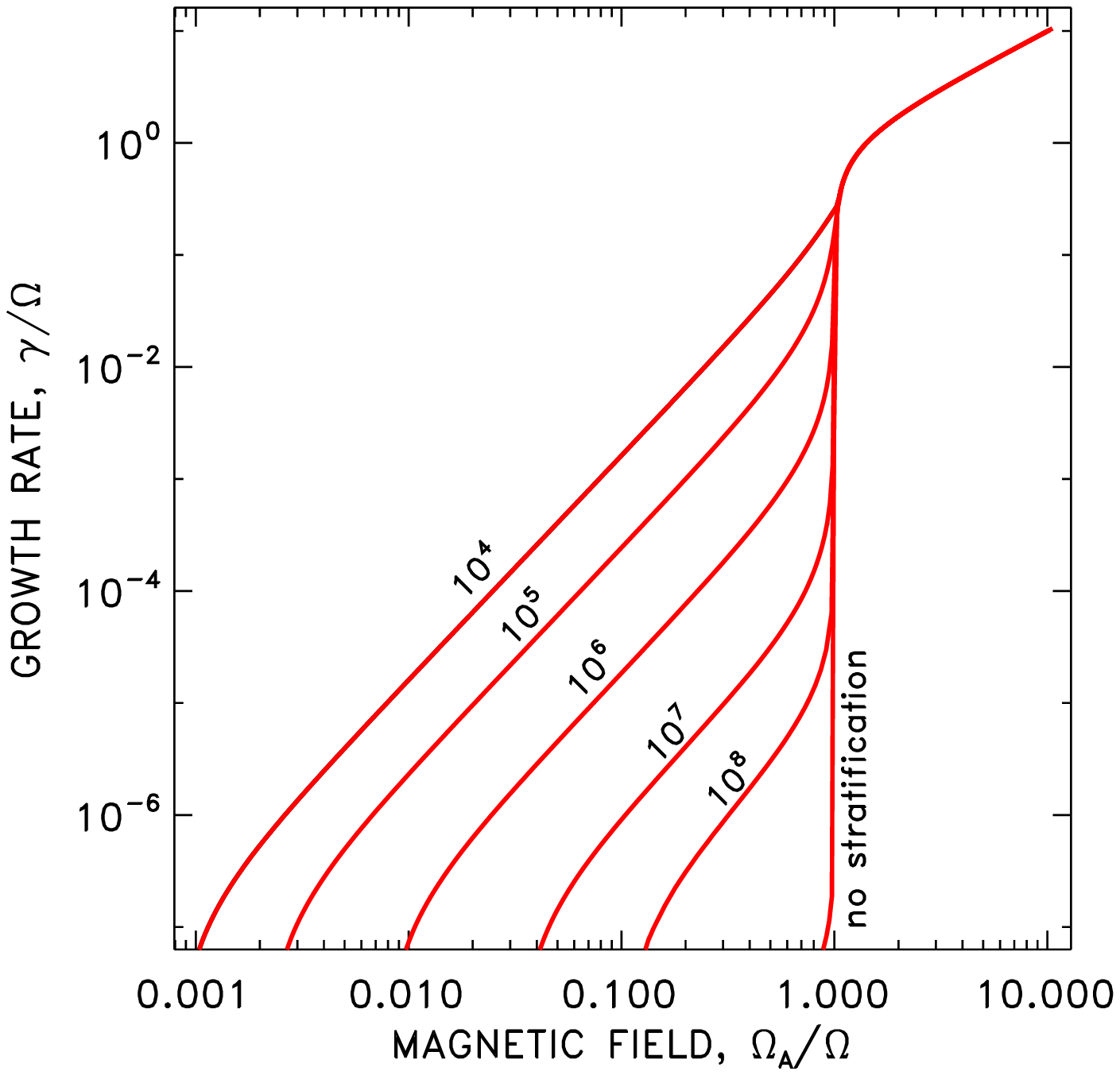}}}
\caption{\label{f1} Growth rates of nonaxisymmetric disturbances of a background field with
    quadrupolar symmetry for small (left) and large (right) thermal
    diffusivity. The lines are marked by $C_\chi$ after equation (\ref{3}).
    The growth rates for weak fields with $\Omega_\mathrm{A} < \Omega$
    initially increase with $C_\chi$ but then decrease for very
    large thermal diffusivities.}
\end{center}
\end{figure}

The growth rates of the  instability for various values of the thermal
diffusivity parameter $C_\chi$ are shown in figure.~\ref{f1}. All the
lines converge for sufficiently strong fields, $\Omega_\mathrm{A}
> \Omega$, showing that then the  instability  is
insensitive to diffusion. The growth rates $\gamma$ for strong fields can
be estimated as $\gamma \simeq \Omega_\mathrm{A}$ \citep{S99}.

The magnetic instability in adiabatic  fluids, i.e.
with $\chi = 0$, is suppressed by the rotation, \citep{PT85,C03}. But for  finite thermal diffusion also weak fields under the presence of
superAlfv\'enic rotation, $\Omega > \Omega_\mathrm{A}$, are unstable
though with small growth rates. The growth rates initially increase with increasing $C_\chi$ to saturate for about $C_\chi = 1$ (figure \ref{f1}, left). Due to its very small growth rates the instability of weak fields ($\Omega_\mathrm{A} < \Omega$) for both small and very large $C_\chi$ cannot be detected with time stepping codes (see Braithwaite 2006).

In the wide range
\begin{equation}
  0.1 < C_\chi < 10^4
    \label{411}
\end{equation}
also  a weak-field instability is faintly sensitive to thermal diffusivity. The growth rate runs as
\begin{equation}
  \gamma\propto (\frac{\Omega_{\rm A}}{\Omega}) \ \Omega_{\rm A} .
    \label{42}
\end{equation}
Only for very large $C_\chi$ when the buoyancy is totally suppressed
the results for unstratified fluids with their strong rotational
suppression of the instability are reproduced (figure \ref{f1},
right).

The case of one magnetic belt, $B_\phi \sim \sin\theta$, is
exceptional \citep{PT85}. When the ratio of rotational velocity to
Alfv\'en velocity is uniform then there is no preferred latitude for
the instability to start. Finite thermal diffusion promotes the
instability as the effect of diffusion depends on the latitudinal
scale of the disturbances. The isothermal limit
$\chi\rightarrow\infty$ represents also  the case of unstratified
fluid and the instability is totally suppressed for fast rotation.
 \subsection{Dipolar magnetic geometry}
\begin{figure}
\begin{center}{
\resizebox*{8cm}{!}{\includegraphics{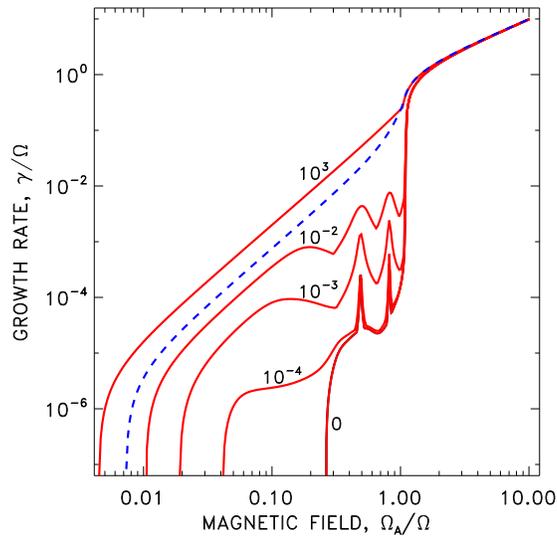}}
}
\caption{\label{f2}
Growth rates of the instability for two magnetic belts of opposite
    signs ($n=1$). The lines are marked with the corresponding
    $C_\chi$-parameter (\ref{3}). The dashed line shows the results
    for $g=0$ and/or $\chi\to \infty$.}
\end{center}
\end{figure}

We find a  different situation for dipolar background fields ($n=1$)
with two belts of opposite polarity as  resulting from the
interaction  of differential rotation and dipolar poloidal fields.
Figure~\ref{f2} shows the results. In the above model of quadrupolar
fields, the vertical lines of figure~\ref{f0} for adiabatic
($\chi=0$) and isothermal ($\chi \to \infty$) disturbances are
rather parallel and close together. Now  the (dashed) line for $\chi
\to \infty$ (or, what is the same, for unstratified fluids) and the
vertical line for $\chi=0$ differ completely. The dashed line is
close below the line for $C_\chi$= 1000 for which the maximum growth
rates appear. It also shows the typical behavior (\ref{42}) for
subequipartition fields ($\Omega_\mathrm{A} < \Omega$).
Consequently, one finds the relation (\ref{42}) true for all
nonadiabatic disturbances in fluids with $C_\chi> 0.1$. Hence, if
the very small growth rates are acceptable, then already toroidal
fields with $\Omega_\mathrm{A} \simeq 0.01\ \Omega$ become unstable
if the $C_\chi$ is not too small. For the Sun the maximum strength
of stable fields is about 500 Gauss (cf. Spruit 1999). Note that the
growth rates for the solar value of $C_\chi= 0.1$ differ slightly
from the growth rates for isothermal case. For the stellar magnetic
fields with dipolar parity, $C_\chi\simeq 0.1$ already represents
the situation for $\chi \to \infty$. The inclusion of the buoyancy
is thus not even necessary. This statement does not hold for the
above model of quadrupolar field geometry.

\begin{figure}
\begin{center}{
\resizebox*{5.3cm}{!}{\includegraphics{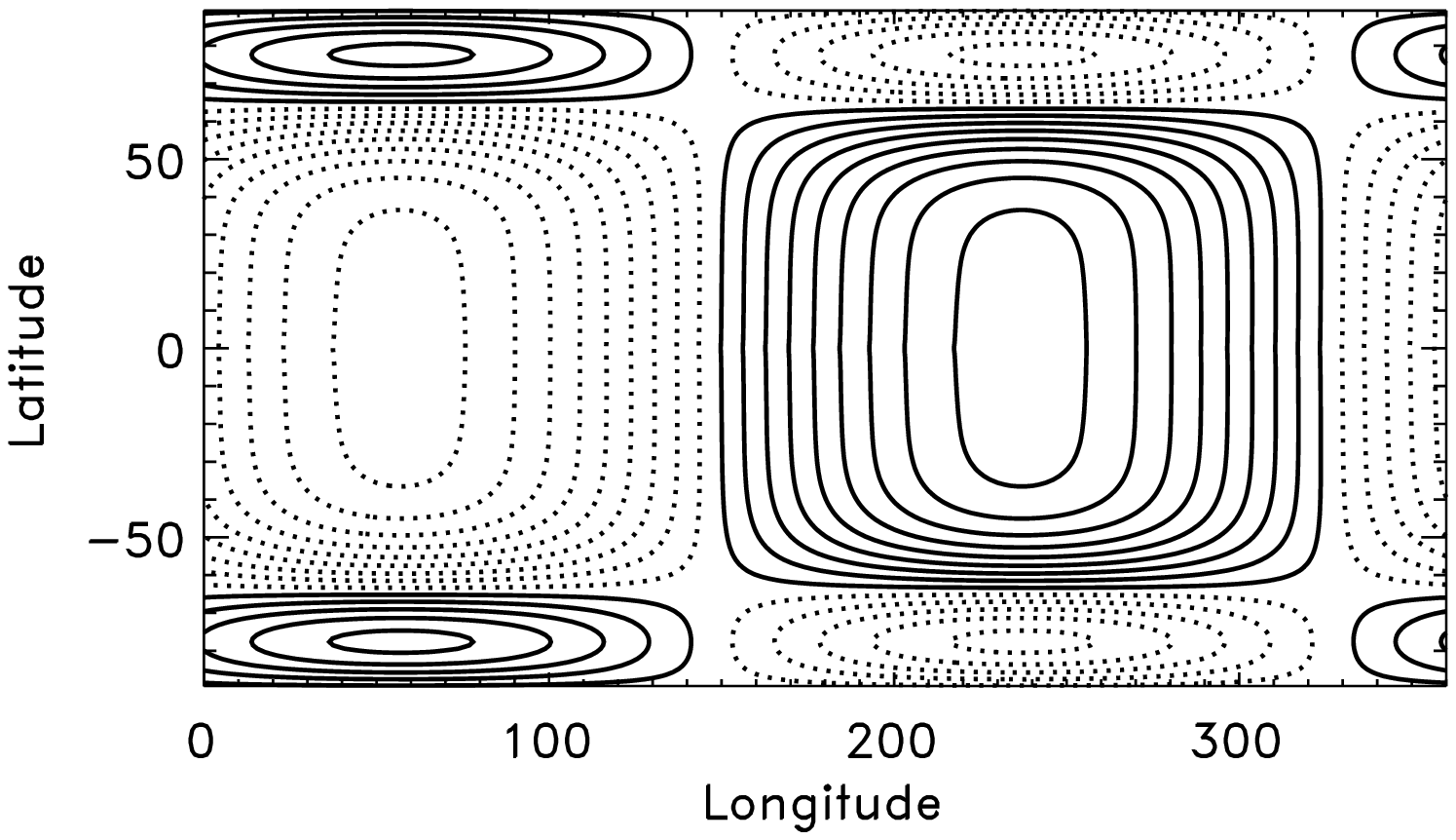}}
\hspace{0.2truecm}
\resizebox*{5.3cm}{!}{\includegraphics{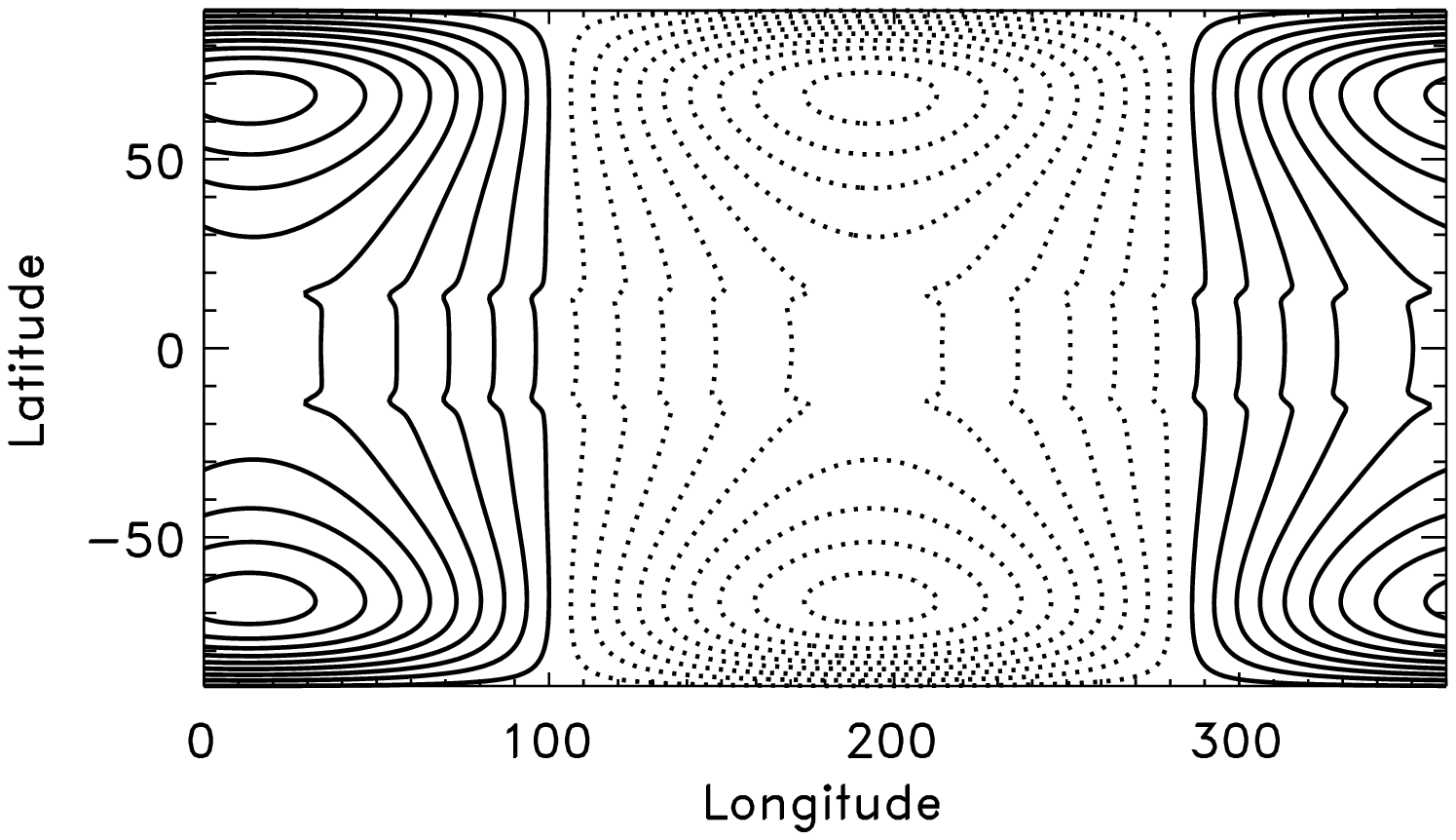}}
\hspace{0.2truecm}
\resizebox*{5.3cm}{!}{\includegraphics{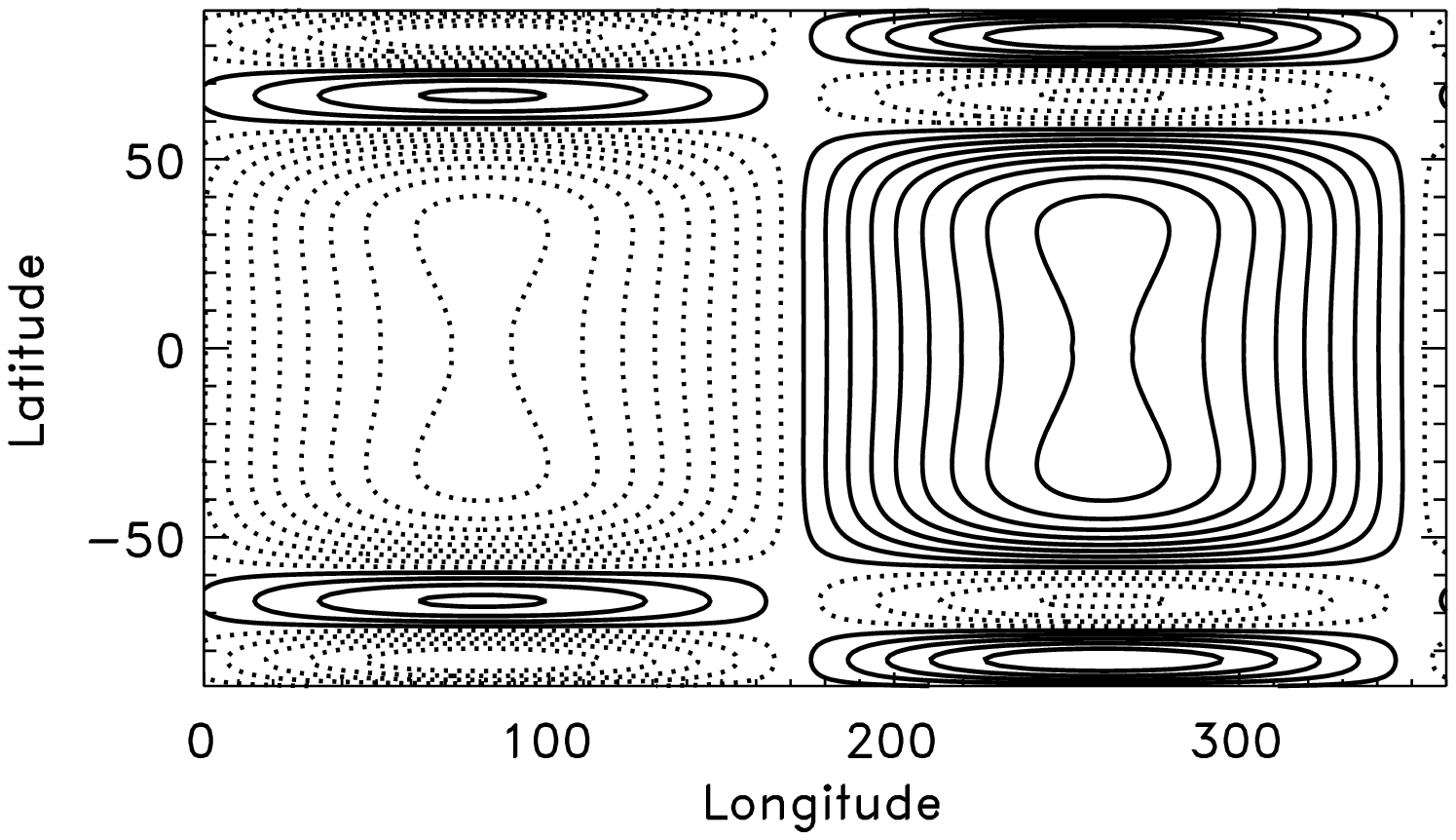}}}
\caption{\label{f3}
Streamlines of the toroidal flow for the most rapidly growing
    eigenmodes for the small value $C_\chi = 10^{-4}$. The left panel corresponds
    to the left peak in the growth rates of figure~\ref{f2}.
    The right panel is for the right peak and the middle panel is for a
    nonresonant mode between the peaks. Full and dotted lines show
    opposite senses of circulation.}
\end{center}
\end{figure}

A very new feature of the considered two-belts geometry  are
resonances. The lines with fixed but small  $C_\chi$ show two peaks
at magnetic field amplitudes slightly below the equipartition level
of $\Omega_{\rm A} = \Omega$. The resonant eigenmodes do not appear
in the  one-belt model and they show a more detailed fine-structure
than the nonresonant ones (figure~\ref{f3}). As a doubling of the
resolution does  not change the results the numerically detected
resonances seem to be  real. Note that all the unstable modes are
global in horizontal dimensions.

\begin{figure}
\begin{center}{
\resizebox*{8cm}{!}{\includegraphics{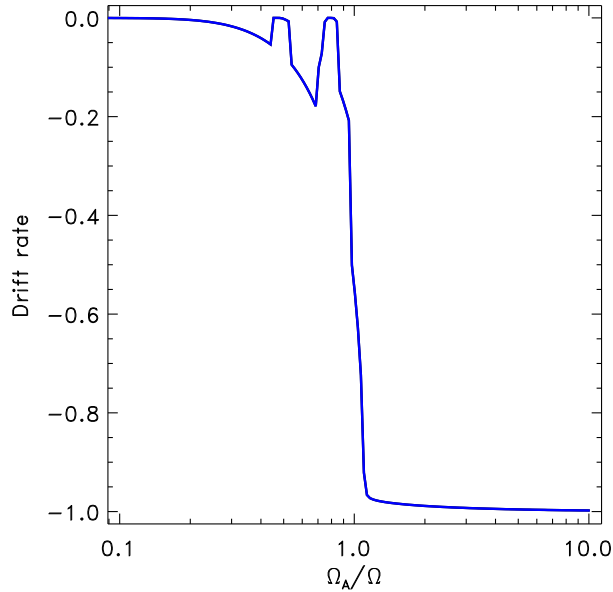}}
}
\caption{\label{f4}
Drift rates of the most rapidly growing modes in the corotating reference
    frame. The rates are normalized with $\Omega$. Negative values mean the
    counter-rotation drift. Resonant modes corotate with the fluid.
    The eigenmodes for strong fields, $\Omega_\mathrm{A} > \Omega$,
    are resting in non-rotating frame. The plot is for $C_\chi = 10^{-4}$.}
\end{center}
\end{figure}

The resonant nature of the peaks in figure~\ref{f2} is illustrated
by figure~\ref{f4} showing the drift velocities for a small $C_\chi
= 10^{-4}$. Generally, the unstable modes drift against the
direction of the global rotation. Note that the drift rates of the
resonant modes are zero. These  modes corotate with the fluid.

For strong fields with $\Omega_\mathrm{A} >\Omega$ the instability
pattern  does not follow the basic rotation. The normalized drift
rates approach the value of -1  which means that the modes are
resting in the inertial frame of reference. Test calculations showed
that this result also holds for $\rm Pm=1$ which for the interior of
hot stars is not unrealistic due to the large radiative viscosity.
The nonaxisymmetric field pattern produced by instability of a
strong field seems to rest in the laboratory system. If only such a
magnetic pattern is observed on a star then it seems to rest or to
exhibit an extremely slow rotation. The steepness of the drift rate
profile of figure~\ref{f4} suggests that the transition between the
nondrifting solution (where the difference between the rotation of
the star and the rotation of the magnetic field is very small) and
the drifting solution (where the magnetic pattern rotation
disappear) is very sharp. If Ap stars are assumed as stars with
unstable toroidal background fields (which themselves are invisible)
then two groups among them should exist depending on the ratio of
$\Omega_\mathrm{A}$ to  $\Omega$. The magnetic field pattern of the
group with  $\Omega_\mathrm{A} < \Omega$ rotates slightly slower
than the star but the magnetic field pattern of the group with
$\Omega_\mathrm{A} > \Omega$ should rotate extremely slow.

The instability for strong  fields is very fast. The growth times
for this case after the equation (\ref{eq3})  are shorter than the
rotation period. On the other hand, after (\ref{42}) the instability
of so weak fields that $\Omega_\mathrm{A} < \Omega$ is much  slower.
It is, however,  hardly controlled by the thermal diffusion. The
instability of weak fields in stellar radiation zones should thus
not be too sensitive to the chemical details.
\section{Differential rotation}
Sofar we have assumed the  stellar rotation as rigid. This is only true if the star is old enough. A possible differential rotation produces a strong toroidal field from the original   fossil poloidal field. The resulting Maxwell stress suppresses the differential rotation producing an almost rigid rotation after the Alfv\'en travel time estimated for the poloidal field which for hot stars with fields of order mGauss lasts longer than 10 Myr.

Hence, for young stars the instability of the field pattern must be considered under the presence of differential rotation. As the current-driven instability is basically nonaxisymmetric one must expect the action of the differential rotation as stabilizing so that a possible instability might occur only after the Alfv\'en  travel time.
\begin{figure}
\begin{center}{
\resizebox*{8cm}{!}{\includegraphics{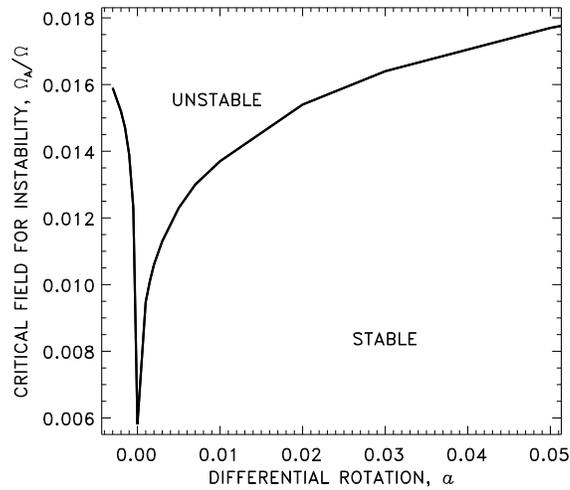}} }
\caption{\label{f7} The stabilizing influence of differential
rotation on toroidal  fields with dipolar symmetry. The critical
magnetic fields for onset of the instability optimized by  choice of the wave number are shown. Note that
the stabilization by differential rotation does only slightly depend
on the sign of rotational shear. $C_\chi=0.1$}
\end{center}
\end{figure}

We have worked with the simple rotation law
\begin{equation}
 \Omega = \Omega_0 (1 + a' r^2 \sin^2 \theta),
    \label{45}
\end{equation}
which in its simplified local formulation reads as $\Omega \propto  1 - a\ \cos^2 \theta$. Equation (\ref{45})
describes a rotation law with cylindric isolines. For negative  $a$ the rotation rate decreases  outwards and v.v. The stabilizing effect of the differential rotation should  {\em not} depend on the sign of $a$.

Here we only consider the magnetic field geometry (\ref{4}) for dipolar field structures, i.e. with antisymmetry of the fields with respect to the equator. Figure \ref{f7}  gives the results for $C_\chi=0.1$. The critical magnetic frequencies  are optimized by choice of  the wave numbers. We find the stabilization by differential
rotation as  highly effective. Already very small shear values lead to an increase of the critical
magnetic field by a factor of (say) five. The influences of both the sign and the real value of $a$ are  small.

The figure \ref{f7} suggests a very efficient stabilization of the toroidal fields by any kind of differential rotation if the  field is antisymmetric with respect to the equator. The same might  {\em not} be true for other field geometries (see R\"udiger and Schultz 2010). Hence, we find the toroidal magnetic fields in stellar tachoclines with their strong differential rotation (if  due to a fossil poloidal dipolar field) much more stable than they are in the rigidly rotating cores of stars.

The stabilizing action of differential rotation does hardly depend on the form of the rotation law. If the star rotates nonhomogeneously then higher amplitudes of the induced toroidal fields remain stable. The results of this section suggest the importance of further studies of the interaction of magnetic fields and differential rotation. If the fields are produced by a dynamo mechanism then the magnetic field geometry can easily differ (like in galaxies) from that considered in the present paper.

\section{Stellar magnetism}
\begin{figure}
\begin{center}{
\resizebox*{7cm}{!}{\includegraphics{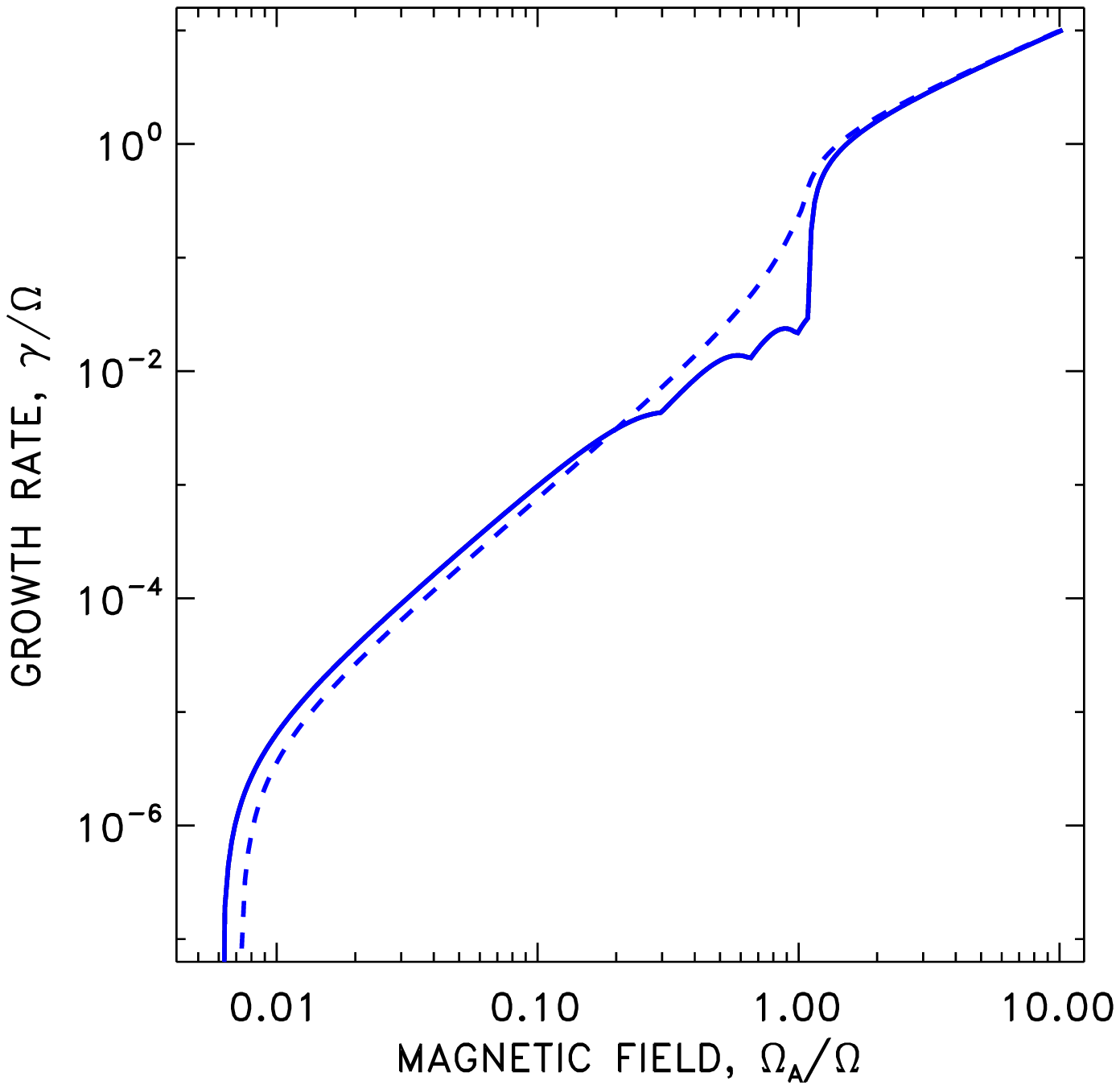}}
\hspace{0.3truecm}
\resizebox*{7cm}{!}{\includegraphics{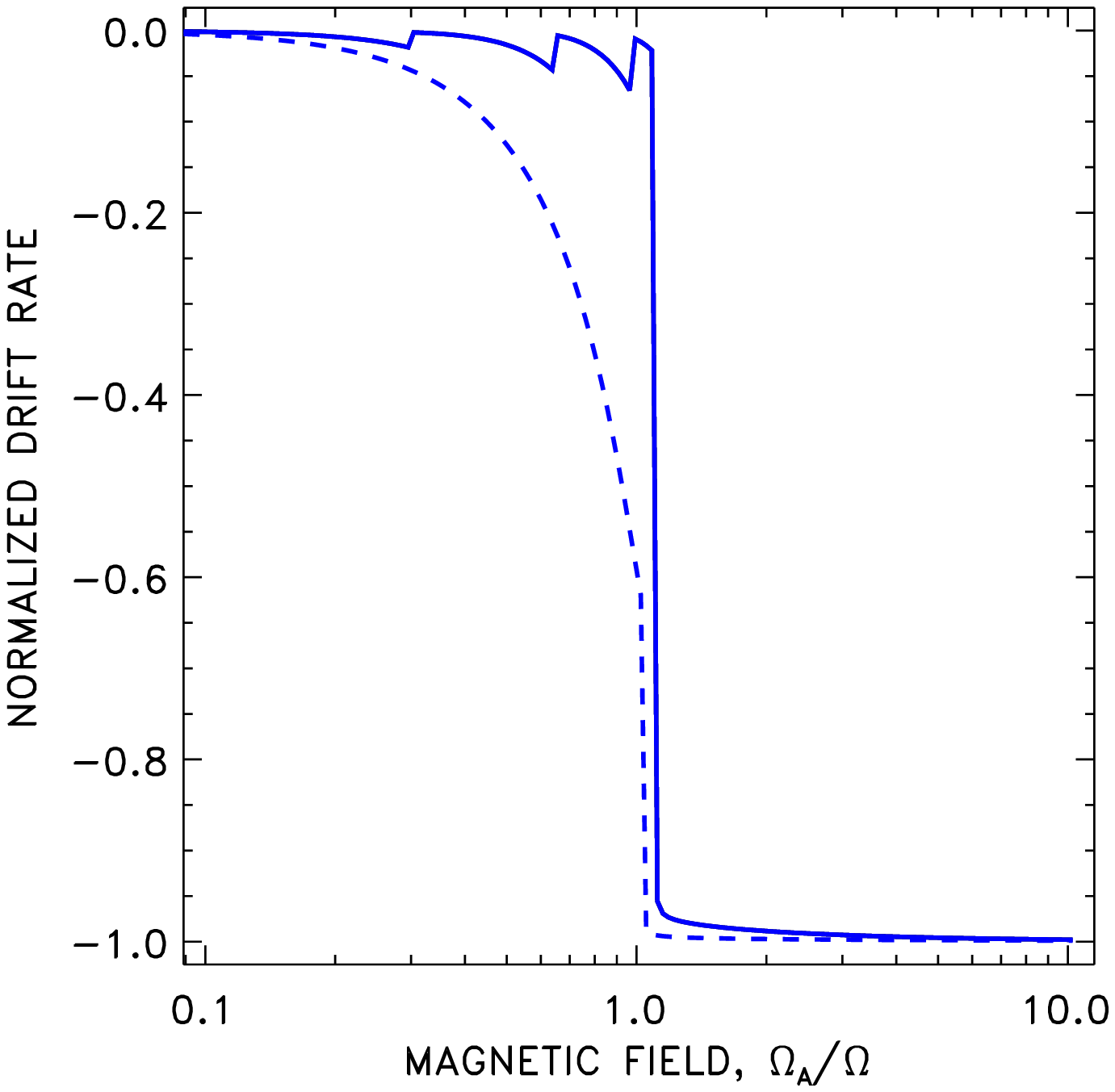}}
}
\caption{\label{f6}
Dipolar field geometry:  Normalized growth rate (left) and drift
    frequency (right) for the solar value of $C_\chi = 0.1$ (solid lines).
    The dashed  lines are for $g=0$ and/or $\chi\to \infty$. Note the small
    differences between solid and dashed lines and the abrupt changes of the
    eigenfrequencies at $\Omega_\mathrm{A} = \Omega$. Test calculations
    have shown that the location of the footpoints of both the dashed and
    the solid lines do basically not depend on the magnetic Prandtl
    number $\nu/\eta$.}
\end{center}
\end{figure}

We have shown that for realistic values of the heat diffusivity the
growth rates of the kink-type instability in stellar radiation zones
do hardly  differ from the growth rates obtained for fully
incompressible models without buoyancy. There are, however, strong
differences for other types of the magnetic geometry. Cylindric
models with uniform Alfv\'en frequency do not completely cover the
situation for spherical models with toroidal fields of dipolar
parity.  For stellar applications the main results for the
instability of such fields are given in figure~\ref{f6}. Note how
well the approximation without gravity $g=0$ and/or $\chi\to \infty$
(dashed line) works in comparison to the `exact' profile for $C_\chi
= 0.1$ (solid line).

After the figure~\ref{f6} (left) three groups of hot stars can be
distinguished in dependence on the amplitude of their toroidal
fields. The toroidal magnetic field of the first group fulfills the
relation $\Omega_\mathrm{A} < 0.01\ \Omega$ so that it  remains
stable. If the field amplitude  exceeds the lower limit (or the
rotation is slow enough) then there are two possibilities. If it is
not too strong, i.e. $\Omega_\mathrm{A} < \Omega$, then it becomes
unstable with very small rates of growth and azimuthal drift.

If strong enough, the poloidal component of the resulting  nonaxisymmetric field  should be observable. The critical Alfv\'en velocity for a typical hot star is about 10 km/s corresponding to a magnetic field of order 10$^6$ Gauss. If only 1\% of the magnetic energy move to the poloidal perturbation, the amplitude of poloidal field is about 10$^5$ Gauss (see Gellert {\it et al.} 2007). This value is even larger than the observed fields of Ap stars.

The third group of stars fulfills the condition $\Omega_\mathrm{A} > \Omega$. They are unstable with very short growth times. Their drift rates, however, approach the value of -1 so that the magnetic patterns may only show a very slow global rotation. There are indeed examples among the group of the Ap stars with rotation periods of several years. The bright star $\gamma$ Equ has a rotation period longer than 70 yr.

In the light of the presented theory  the basic fact that the Ap stars are slow rotators compared with the normal A stars mainly means that slow rotation is less stabilizing for the toroidal magnetic fields. With other words, the condition  $\Omega_\mathrm{A} > 0.01\ \Omega$ for instability is more easily fulfilled for slow rotators. It is thus understandable with our results that for slow rotation weaker toroidal fields become unstable and also the resulting amplitude of the $m=1$ mode is smaller for slow rotation than for fast rotation -- which indeed is observed \citep{Hea07}.

The condition $\Omega_\mathrm{A} > 0.01\ \Omega$ for instability
could easily be the counterpart of the lower limit of about 300
Gauss found by Auri\'ere {\it et al.} (2007) for magnetic fields of
Ap stars. The existence of rather strong toroidal magnetic fields
within stellar radiation zones can be thought of as the outcome of
the differential rotation and a weak fossil poloidal field. For the
above calculations we have assumed that the differential rotation
only exists before the magnetic instability develops. The reason is
that  (any form of)  differential rotation  stabilizes the
instability of fields with equatorial antisymmetry. From this point
of view all the Ap stars are considered as  (slow) rigid rotators.

A complete explanation of the Ap star magnetism still meets open questions. So the axis of the magnetic field pattern is obviously {\em not} orthogonal to the axis of rotation (Oetken 1977). The obliquity of the field, i.e. the ratio of nonaxisymmetric and axisymmetric field parts, depends on the rotation rate: it is maximum for large $\Omega$ (Landstreet and Mathys 2000). The instability of a single  $m=1$ mode cannot explain this finding. It is also known that the magnetic Ap stars do {\em not} exist  close to the ZAMS, they are concentrated toward the center of the main-sequence. The earliest observed evolution time across the main-sequence  of a magnetic Ap and Bp  star is about 20 Myr, no one younger magnetic star has been observed (Hubrig {\it et al.} 2000). There are also possibilities to explain these empirical findings but those are  beyond the scope of the present paper.

{\sl Acknowledgements.} This work was supported by the Deutsche
Forschungsgemeinschaft and by the Russian Foundation for
Basic Research (project 09-02-91338).


\label{lastpage}

\end{document}